\documentclass[pdflatex,sn-mathphys-num]{sn-jnl}

\usepackage{graphicx}%
\usepackage{multirow}%
\usepackage{amsmath,amssymb,amsfonts}%
\usepackage{amsthm}%
\usepackage{mathrsfs}%
\usepackage[title]{appendix}%
\usepackage{xcolor}%
\usepackage{textcomp}%
\usepackage{manyfoot}%
\usepackage{booktabs}%
\usepackage{algorithm}%
\usepackage{algorithmicx}%
\usepackage{algpseudocode}%
\usepackage{listings}%
\usepackage{macros}

\theoremstyle{thmstyleone}%
\theoremstyle{thmstyletwo}%

\theoremstyle{thmstylethree}%

\begin{document}

\title[]{\ours: Investigating Opportunities to Enhance Localization and Reconstruction in Image-based Arthroscopy Navigation via External Cameras}


\author*[1]{\fnm{Hongchao} \sur{Shu}} \email{hshu4@jhu.edu}

\author[1]{\fnm{Lalithkumar} \sur{Seenivasan}}
\author[1]{\fnm{Mingxu} \sur{Liu}}
\author[1]{\fnm{Yunseo} \sur{Hwang}}
\author[1]{\fnm{Yu-Chun} \sur{Ku}}
\author[3]{\fnm{Jonathan} \sur{Knopf}}
\author[2]{\fnm{Alejandro} \sur{Martin-Gomez}}
\author[1]{\fnm{Mehran} \sur{ Armand}}

\author*[1]{\fnm{Mathias} \sur{Unberath}}\email{unberath@jhu.edu}




\affil*[1]{\orgname{Johns Hopkins Univeristy}, \orgaddress{\city{Baltimore}, \state{Maryland}, \country{USA}}}

\affil[2]{\orgname{University of Arkansas}, \orgaddress{\city{Fayetteville}, \state{Arkansas}, \country{USA}}}

\affil[3]{\orgname{Arthrex Inc.}, \orgaddress{\city{Naples}, \state{Florida}, \country{USA}}}

\maketitle

\section{Introduction}\label{sec:intro}

Arthroscopy is a minimally invasive surgery (MIS) used to diagnose and treat intra-articular structures. However, limited visualization introduces challenges such as reduced spatial awareness, depth perception, and a narrow field of view~\cite{bartoli2012computer}. Although navigation systems, enabled by optical and electromagnetic (EM) trackers, address these limitations, they impose workflow constraints due to line-of-sight requirements and sensitivity to metallic interference~\cite{samarakkody2016use}. Vision-based SLAM provides an alternative navigation solution that imposes fewer restrictions on the workflow but suffers from scale ambiguity and drift~\cite{shu2024seamless}. 
To overcome these challenges, we introduce the \ours system, the first to enable purely vision-based intraoperative arthroscopy navigation. Unlike existing methods, it achieves arthroscope localization and intra-articular reconstruction without relying on expensive or bulky hardware. By combining monocular endoscopic tracking~\cite{teufel2024oneslam} with visual odometry~\cite{9440682} using an external camera, \ours provides accurate localization and reconstruction at a significantly lower cost, offering a more accessible solution for surgical navigation.


Our key contributions include the development of the~\ours system, a low-cost and portable solution for intra-operative arthroscopy navigation. Unlike existing methods,~\ours enables real-time arthroscope localization and detailed articular region reconstruction solely from visual data. We further demonstrate its potential for intra-operative guidance by quantitatively evaluating its tracking and reconstruction accuracy, highlighting its effectiveness and distinct advantages.

\section{System Overview}\label{sec:methods}

\begin{figure}[!t]
    \centering
    \includegraphics[width=\textwidth]{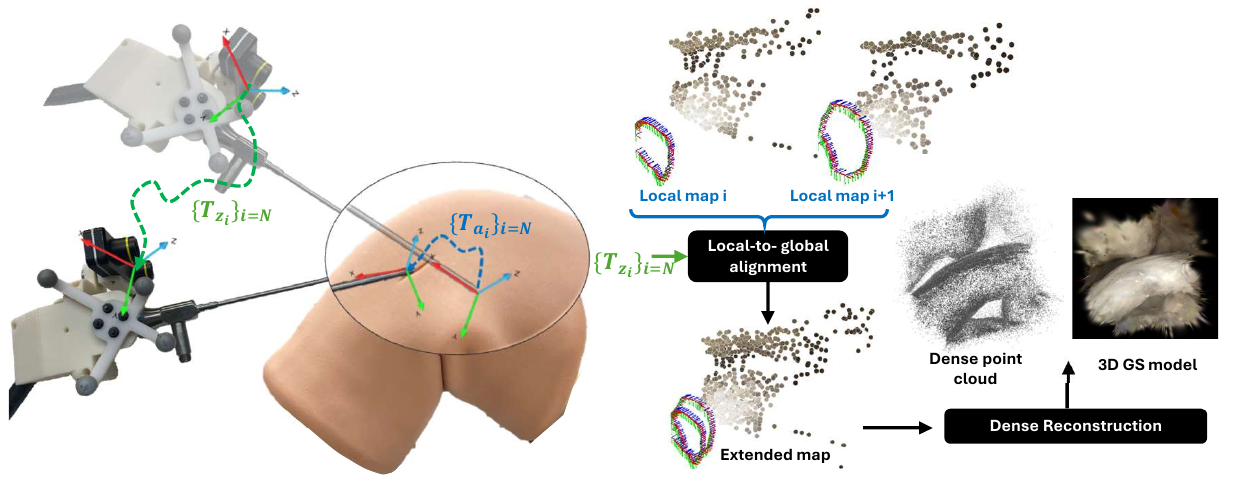}
    \caption{\ours integrates an endoscope, stereo camera, and 3D Gaussian Splatting for dense intra-articular 3D reconstruction. We use TAP models for pose estimation and sparse reconstruction, followed by dense reconstruction using pseudo depth and 3D GS training. Multi-camera calibration aligns the endoscope scene with the external camera scene, establishing absolute scale and spatial relationships.}
    \label{fig:main}
\end{figure}

Our \ours system (Fig. ~\ref{fig:main}) integrates an arthroscope with a rigidly mounted external stereo camera and leverages vision-based localization and 3D Gaussian Splatting (3D GS)~\cite{shu2023twin} for navigation and dense 3D reconstruction of intra-articular structures. Firstly, we use a custom 3D-printed mount that rigidly attaches the external camera to the arthroscope. A RANSAC-based image selection method is then employed to calibrate and obtain the intrinsic parameters of both cameras. Furthermore, a hand-eye calibration~\cite{Furrer2017FSR} aligns the rigid camera setup, with offset compensation~\cite{ma2020} mitigating positional errors along the arthroscope shaft. Secondly, we employ our vision-based tracking and reconstruction pipeline~\cite{shu2024seamless} for local scene reconstruction, and ORBSLAM ~\cite{9440682} for external camera external localization. We introduce a local-to-global registration module that corrects scale ambiguity in monocular arthroscopic vision by finding an optimal transformation that maps local reconstructions to a consistent global model, ensuring smooth integration of scene data.

\section{Experiment and Results}\label{sec:exp}
 We assessed the calibration accuracy of the \ours system using Reprojection Error (RPE) to evaluate both intrinsic and hand-eye calibration. The external camera demonstrated an RPE of 0.98 pixels, while the arthroscope achieved 2.90 pixels. Hand-eye calibration yielded an RPE of 16.00 pixels for the external camera and 47.43 pixels for the arthroscope, translating to approximately 0.40 mm and 1.90 mm, respectively. This higher RPE for the arthroscope, driven by factors like lens distortion and the long shaft, highlights the need for vision-based localization in arthroscopy.
 


We evaluated tracking accuracy by comparing the system's camera trajectories with ground truth from optical tracking (Table.~\ref{tab:track}). We used the absolute trajectory error (ATE) to assess global alignment and the relative trajectory error (RTE) to determine motion consistency. We used the root-mean-square (RMS) of translation acceleration to quantify the smoothness of trajectories. The results showed reasonable RTE ($Trans (mm) = 1.09, Rot (^\circ) = 64.47$) and ATE values ($Trans (mm) = 1.08, Rot (^\circ) = 1.76$) on average, particularly in translation, with smoother predicted trajectories ($RMS=7.2\times10^{-4}$) than ground truth ($RMS=1.3\times10^{-4}$), as indicated by the acceleration values of RMS on average. This demonstrates the effectiveness of our alignment in mitigating the impact of suboptimal calibration.




For evaluating reconstruction (Table.~\ref{tab:reconstruction}), we aligned the local maps with the global map to generate dense 3D GS models. Target Registration Error (TRE) was calculated as root mean square error (RMSE) between reconstructed point maps and CT scans, yielding an average RMSE of $2.16$ mm and an average Hausdorff distance of $10.28$ mm. This indicates that the reconstructed surfaces are mostly aligned with the ground truth except for a small region of poorly aligned points. The high PSNR ($22.19$) and SSIM ($0.69$) values confirmed the visual fidelity of the model to the real world structures, validating the feasibility of \ours for intraoperative guidance.

\begin{table*}[!t]
\caption{Average tracking accuracy with ATE, RTE as the trajectory absolute trajectory error, smoothness reflects how well the system maintains fluid motion.}
\centering
\label{tab:track}
\scalebox{0.55}{
\begin{tabular}{c|cc|cc|cccc}
\toprule
\multirow{2}{*}{\textbf{Trials}} & \multicolumn{2}{c|}{\textbf{ATE}} & \multicolumn{2}{c|}{\textbf{RTE}} & \multicolumn{4}{c}{\textbf{Smoothness}}               \\
                        & Trans (mm)    & Rot ($^\circ$)    & Trans (mm)   & Rot ($^\circ$)   & GT $RMS (\vec{a})$ & GT $RMS (\alpha)$ & Pred $RMS (\vec{a})$ & Pred $RMS (\alpha)$ \\
\midrule
A   & 0.51& 23.83& 0.41& 1.16& $1.\times10^{-3}$& $7.5\times10^{-3}$ & \textbf{$4.2\times10^{-5}$}& \textbf{$2.8\times10^{-3}$}\\
B   & 1.12& 29.78& 0.73& 3.62& $1.\times10^{-3}$& $7.5\times10^{-3}$ & \textbf{$2.03\times10^{-5}$}& \textbf{$3.4\times10^{-3}$}\\
C   & 0.74& 41.02& 0.53& 1.86& $6.9\times10^{-4}$& $5.9\times10^-3$ & \textbf{$4.22\times10^{-5}$}& \textbf{$2.4\times{10^{-3}}$}\\
D   & 2.50& 154.23& 2.60& 1.76& \textbf{$5.8\times{10^{-4}}$}& \textbf{$5.5\times{10^{-3}}$}& \textbf{$3.6\times{10^{-4}}$}& \textbf{$3.3\times{10^{-3}}$} \\
E   & 1.23& 121.23& 1.96& 1.54& {$5.3\times{10^{-4}}$}& {$4.9\times{10^{-4}}$}& {$3.1\times{10^{-4}}$}& {$3.6\times{10^{-3}}$} \\
F   & 0.49& 16.74& 0.3& 0.62& $5.2\times10^{-4}$& $5.5\times10^{-3}$ & \textbf{$2.44\times10^{-5}$}& \textbf{$1.7\times10^{-3}$}\\
\bottomrule
\end{tabular}
}
\footnotetext[1]{0}
\footnotetext[2]{0}
\end{table*}


\begin{table*}[!t]%
\caption{3D Reconstruction Results showing RMSE and Hausdorff distance for geometry quality, and PSNR and SSIM for rendering fidelity.}
\centering
\label{tab:reconstruction}
\scalebox{0.6}{
\begin{tabular*}{\textwidth}{@{\extracolsep\fill}llllll@{\extracolsep\fill}}%
\toprule
\multirow{2}{*}{\textbf{Trials}} & \multirow{1}{*}{\textbf{RMSE}}&\textbf{Hausdorff} & \multirow{2}{*}{\textbf{PSNR}} & \multirow{2}{*}{\textbf{SSIM}} \\
& \textbf{(mm)} & \textbf{Distance (mm)} & & \\
\midrule
\multirow{1}{*}{A} &      3.52 &   8.67 &     23.21 &   0.74\\
\multirow{1}{*}{B}  &    2.23 &  13.32 &     19.23 &    0.63 \\
\multirow{1}{*}{C}  &    2.39 & 11.70 &     22.71 & 0.66 \\
\multirow{1}{*}{D}  & 1.75 &  9.70 &  26.83 & 0.63 \\
\multirow{1}{*}{E}  &1.87 &  8.91 &  21.47 & 0.73 \\
\multirow{1}{*}{F} & 1.2 &  9.40 &  19.70 & 0.80 \\
\bottomrule
\end{tabular*}}
\end{table*}

\section{Conclusion and Future Works}

In this work, we present the \ours system, an arthroscopy navigation and reconstruction solution addressing challenges like scale ambiguity, tracking drift, and sensitivity to rapid movements in monocular vision-based SLAM. By integrating a multi-camera setup and developing an alignment method to recover scale and reduce calibration inaccuracies, the system achieves smoother tracking and larger-area reconstructions. Future work will benchmark the system on ex vivo specimens to validate its effectiveness in diverse surgical settings.

\backmatter

\bmhead{Acknowledgements}
This work was conducted under a sponsored research agreement with Arthrex Inc. and supported by internal funds from Johns Hopkins University.

\bibliography{sn-bibliography}

\end{document}